# Quantum Advantage of Threshold Changeable Secret Sharing Scheme

Xiaogang CHENG[1], Ren GUO[2], Changli ZHOU[1]

1. College of Computer Science and Technology, Huaqiao University, Xiamen 361021, China
2. College of Business Administration, Huaqiao University, Quanzhou 362021, China

*Abstract:* In TCSS (Threshold Changeable Secret Sharing) scheme, the threshold can be changed to deal with share leakage in the long term. But in classical TCSS, there is no guarantee that old shares are deleted even if the participated parties are honest. So, the changed threshold may not prevent an adversary from reconstructing the secret by the old shares and old threshold number of parties. We show how to tackle this problem quantum mechanically. I.e., quantum mechanically we can make the changed threshold mandatory. So, there is quantum advantage of quantum TCSS over classical TCSS.

## 1. Introduction

By SS (Secret Sharing) scheme, a dealer can distribute a secret to several parties, and only some predefined subset (the so-called access structure) of the parties together can recover the secret. Otherwise, the secret cannot be recovered.

Shamir [1] presented the first and most important $(t,n)$ threshold secret sharing scheme based on polynomial and Lagrange interpolation. Also, there are many other constructions such as secret sharing based on CRT (Chinese Reminder Theorem) [2,3], SS based on hyperplane geometry [4] etc.

There are also many extensions of SS. VSS (Verifiable SS) [5] means that the legitimacy of the secret recovered and the consistency of shares of the parties can be verified. In PSS (Proactive SS) [6,7], the share of the parties can be updated frequently to prevent a eavesdropper to steal the secret in a long period of time. In RSS (Rational SS) [8,9], it is assumed that the participated parties are rational players, who try to maximize their own profits, instead of being honest or malicious in the usual cryptographic setting. Recently there are also many researches on QSS (Quantum SS) [10,11,12,13,14]. In which quantum information (qubits usually) or classical information (classical bits) are distributed and shared quantum mechanically. Quantum rational secret sharing [15,16], which assume that the participating parties are rational, can be fairer and more robust in practice compared to traditional quantum secret sharing schemes. There are also works on combining classic and quantum secret sharing scheme [17], i.e. hybrid secret sharing. And the scheme in [17] also has the advantage of requiring less quantum bit than existing schemes.

In threshold secret sharing scheme, usually it is only required that no less than the threshold number of parties should be present to reconstruct the secret. Other than that, more parties will not bring any advantage. While in [14], more parties will bring communication advantage, i.e., more parties will decrease the communication overhead when reconstructing the secret. In Shamir's $(t,n)$ threshold secret sharing scheme, the threshold $t$ is fixed and cannot be changed. While in TCSS (Threshold Changeable SS) [18,19,20,21], the threshold can be increased to make it harder for an adversary to reconstruct the secret if she/he has stolen some number of shares. But in classical TCSS schemes, the changeable threshold is negotiated by the reconstructing parties and is not mandatory. I.e., when the minimum legitimate number of parties for reconstructing the secret have been corrupted by the adversary, the adversary can easily reconstruct the secret and the changed threshold is void. If TCSS is realized by further splitting and distributing the shares, i.e. each share is further split and distributed to more parties. Then the old share should be deleted, otherwise the secret can be easily reconstructed by those old shares. But in classical computation, there is no guarantee that the old shares are deleted even if the participating parties are honest. For example, an adversary could corrupt a party by hacking into the party's computer, then deleted information could be recovered by the hacker. Or there could be some backup mechanism in the party's computer, the old share is backed up in some other places and later could be found by an adversary and use them to get the secret. So, basically TCSS is theoretically impossible in classical computing environment. In this paper, we show how to overcome this quantum mechanically.

PSS [6] can also enhance the security of SS scheme in a long period time by periodically renewing the share without changing the secret. Of course, to update the share, old shares should be deleted. But again, as we mentioned above, there is no guarantee that the old shares are deleted. And it is still possible for an adversary to recover the old shares and use them to reconstruct the unchanged secret.

The main problem of classical TCSS and PSS of course is that even the participating parties are honest, it cannot guarantee that the shares are deleted. So theoretically it is impossible in classical secret sharing scheme to change the threshold without change the secret since there is no guarantee that the classical information is deleted and cannot be recovered. Classical information can be easily cloned or copied, while in quantum world, there is a well-known "unclonable" property.

In this paper, we study on how to improve the security of TCSS scheme by quantum information. And show that for TCSS, there is a quantum advantage over classical TCSS by constructing a concrete Q-TCSS scheme with enhanced security. In our Q-TCSS scheme, the changed threshold is mandatory, i.e., with number of parties less than the changed threshold, the secret cannot be reconstructed. Basically, this is because quantum information is unclonable, while classical information can be easily copied many times. Our Q-TCSS construction is based on our previous work on multi-layer quantum SS scheme [12], in which there is a hierarchy structure among the participating parties. We also show that although our quantum TCSS is not perfect, classical SS based on our Q-TCSS scheme can be perfect, i.e., no information is leaked if the number of parties is less than the current threshold.

This paper is organized as follows. We present the definition of our TCSS, and review a classical TCSS scheme and its problems in section 2. In section 3, quantum TCSS construction is presented. The security of the Q-TCSS construction is analyzed in section 4. And conclude in section 5.

## 2. Preliminaries
### 2.1 Definition of TCSS:
**Definition 1.** TCSS is composed of the following operations:
1) **Secret Sharing:** The dealer can divide the secret into $n$ shares, and distribute the $n$ shares

to *n* parties.

    2) **Secret Reconstruction:** It is required that all the *n* parties should be present to reconstruct the secret, otherwise the secret cannot be recovered. I.e., the threshold is *n*.

    3) **Threshold Changeability:** The threshold can be increased to N (N > n) by further splitting the secret and distributing to more parties. After the change, then all the N parties should be present to reconstruct the secret.

    Note than our definition of TCSS is different with those in [21]. There are mainly two different points. Firstly, in [21] there is a minimum number of parties t (t<=n), less than which the secret cannot be reconstructed, while in our definition all the participated parties should be present to reconstruct the secret. Secondly, the changeable threshold *l* in [21] is between *t* and *n*, and *t* and *n* are fixed. While in our definition, the threshold is *n*, which can be increased and changeable.

    As we mentioned in the introduction, the definition and construction of TCSS in [21] cannot prevent the adversary from get the secret by corrupting the minimum number of t parties, since t number of parties can reconstruct the secret. While in out definition, if the number of parties changed from *n* to *n*+1, then *n* number of parties will not be able to reconstruct the secret. In this sense, our definition of TCSS is more secure than those in [21].

    Security requirement of TCSS:

    1) If currently the number of all participating parties is *n*, then all the *n* parties should be present to reconstruct the secret. Namely, if less than *n* parties are present, the secret cannot be reconstructed.

    2) If some parties have further split the secret and distributed the share to more new parties, then all the old number of parties will not be able to reconstruct the secret. (This is trivially false in classical cryptographic setting, since old parties can always reconstruct the secret by old shares.)

**2.2 Review of a classical TCSS scheme**

    In this section, we review a recent TCSS scheme of Lein Harn based on bivariate polynomial and discuss some of its problems. In next section, we will show how to overcome these problems by quantum information techniques.

    In normal (t,n) threshold SS scheme, the threshold t is fixed and not changeable. While in TCSS scheme, the threshold t is changeable. The purpose of TCSS is for enhancing the security of SS scheme. In the long run, the secret sharing parties may be corrupted by a adversary, rendering the shared secret insecure. To tacking this, in TCSS the threshold could be improved to prevent the adversary from getting the shared secret even if he/she have corrupted some participated parties. For example, in (3,7) SS scheme, the threshold is 3, if an adversary have corrupted 2 parties, then he will get the secret after corrupting just one more parties. At this point, by TCSS, the threshold could be improved to 5, then at least 3 more parties are required to corrupted by the adversary for him to get the secret. Compared with the previous scenario, now it much harder for the adversary to get the secret. Hence the security of the SS scheme is enhanced.

    In [21], Lein Harn et.al presented a novel TCSS based on bivariate polynomial. The dealer selects a random modular bivariate polynomial in x and y, with the constant being the secret. Such as

$$F(x,y) = 7 + 2x + 2y + 3xy + 3y^2 + 5xy^2 + 7y^3 + 3xy^3 \bmod 11$$

Here of course the secret is the constant 7. The *i*th party will get a share consist of two univariate polynomial $F(i, y)$ and $F(x, i)$, where i is constant. For example, the first party, i.e. when $i = 1$, will get the two polynomials:

$$F(1, y) = 9 + 5y + 8y^2 + 10y^3 \bmod 11$$
$$F(x, 1) = 19 + 13x = 8 + 2x \bmod 11$$

Similarly the 2nd party will get $F(2, y)$ and $F(x, 2)$, 3rd party $F(3, y)$ and $F(x, 3)$ etc.

Obviously $F(x, 0)$ is a univariate polynomial of degree one with the constant being the secret. Hence any two parties i and j (i ≠ j) can reconstruct the secret by using $F(i, 0)$ and $F(j, 0)$ and Lagrange interpolation. Similarly $F(0, y)$ is a univariate polynomial of degree 3 with the constant being the secret. Then any 4 members can reconstruct the secret by using $F(0, i), F(0, j), F(0, k)$ and $F(0, l)$ through Lagrange interpolation.

And the threshold is changeable between 2 and 4, namely 3 parties can also reconstruct the secret as following. Firstly, each party truncate the polynomial $F(1, y)$ to a degree 2 univariate polynomial $F'(1, y)$:

$$F'(1, y) = 9 + 5y + 8y^2 \bmod 11$$

This has the effect of truncating the original polynomial $F(x, y)$ to:

$$F'(x, y) = 7 + 2x + 2y + 3xy + 3y^2 + 5xy^2 \bmod 11$$
$$F(x, y) = a_{00} + a_{01}x + a_{10}y + a_{11}xy + a_{02}y^2 + a_{12}xy^2 \bmod 11$$

Then the three reconstructing parties calculate $F'(1,1), F'(1,2), F'(2,3), F'(2,4), F'(3,5), F'(3,6)$. Now for the polynomial $F(x, y)$, there are six unknown coefficient with the constant being the secret. And we also have six function values at six different points (1,1), (1,2), ...., (3,6):

$$F'(1,1) = a_{00} + a_{01} + a_{10} + a_{11} + a_{02} + a_{12} = 0$$
$$F'(1,2) = a_{00} + a_{01} + 2a_{10} + 2a_{11} + 4a_{02} + 4a_{12} = 7$$
$$F'(2,3) = a_{00} + 2a_{01} + 3a_{10} + 6a_{11} + 9a_{02} + 7a_{12} = 9$$
$$F'(2,4) = a_{00} + 2a_{01} + 4a_{10} + 8a_{11} + 5a_{02} + 10a_{12} = 9$$
$$F'(3,5) = a_{00} + 3a_{01} + 5a_{10} + 4a_{11} + 3a_{02} + 9a_{12} = 1$$
$$F'(3,6) = a_{00} + 3a_{01} + 6a_{10} + 7a_{11} + 3a_{02} + 9a_{12} = 1$$

These are six linear equations in six variables; hence the coefficient can be easily solved by linear algebra. Hence the secret $a_{00}$ can be recovered.

So, the threshold changeability is based on polynomial truncation. And the changed threshold is not mandatory, it is discussed and agreed by the reconstructing parties, other than a mandatory condition. This is in contrast with the original (t, n) threshold SS, where the threshold *t* is mandatory (i.e., less than *t* members cannot reconstruct the secret). So, in the scenario where an adversary has corrupted the 2 parties in the example above, he/she can easily reconstruct the secret by the polynomial $F(x, 0)$ or by truncating the polynomial $F(i, y)$ to a degree 1 univariate polynomial as described above, and the changed threshold 3 is void.

So, for the purpose of protect the shared secret in a long period of time, during which an adversary can have corrupted some members and stolen some shares, the scheme above failed. And in classical computing, this may even be impossible. Since to change threshold, all the parties have to delete their old shares and get new shares. But in classical computing, these is no guarantee that the information has been deleted for sure. Even after honest deletion, it is still possible that there are some backup mechanisms in the computer. Then an adversary or hacker can recover the delete information and using them to reconstruct the secret.

In the next section we show how to solve this quantum mechanically. I.e., based on quantum information techniques, a Q-TCSS is constructed with the property that the changed threshold is mandatory. Any subset with less than all the members will not be able to reconstruct the shared secret. Especially, old shares of old threshold number of members will also not be able to reconstruct the secret.

## 3. Our quantum TCSS Construction

Our quantum TCSS construction is based on our previous work on multi-layer QSS scheme [13]. In a layered QSS scheme, participating parties are organized in a hierarchy way (Fig. 1). In this section, we show how this layered operation can be used for changing the threshold, with basically no changes to the original construction. So, we review the construction in [13] first.

At the first level, only one party hold the quantum bit
$$a|0\rangle + b|1\rangle$$
At the 2nd level though, three parties share the pure entangled quantum state
$$a|000\rangle + b|111\rangle$$
Any one or two of the 3 parties cannot recover the original quantum bit, only when all 3 parties are present the original secret qubit can be recovered. We will prove this in the next section.

And at the 3rd level, nine parties share the entangled quantum state:
$$a|000,000,000\rangle + b|111,111,111\rangle$$
Similarly, any proper subset of the 9 parties cannot recover the qubit, while it can be easily recovered if all 9 parties are present.

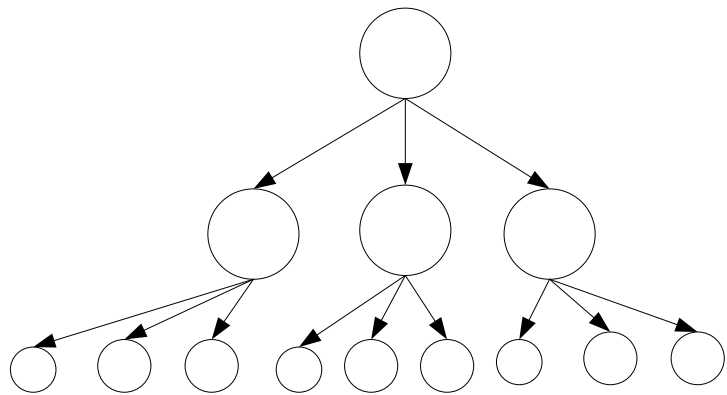

Fig 1. Quantum multi-layer secret sharing scheme

And the quantum circuit for achieving this kind of transformation is as the following:

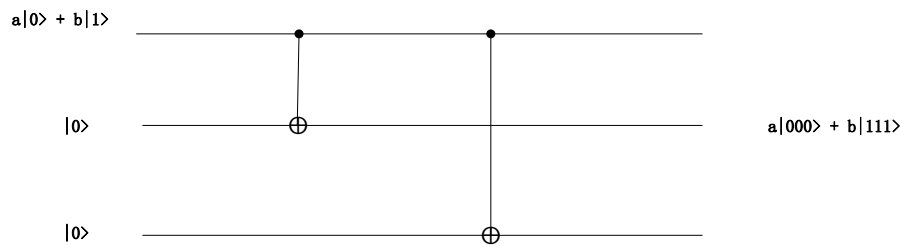

Fig 2. Quantum circuit for the 2$^{nd}$ level of multi-layer quantum secret sharing

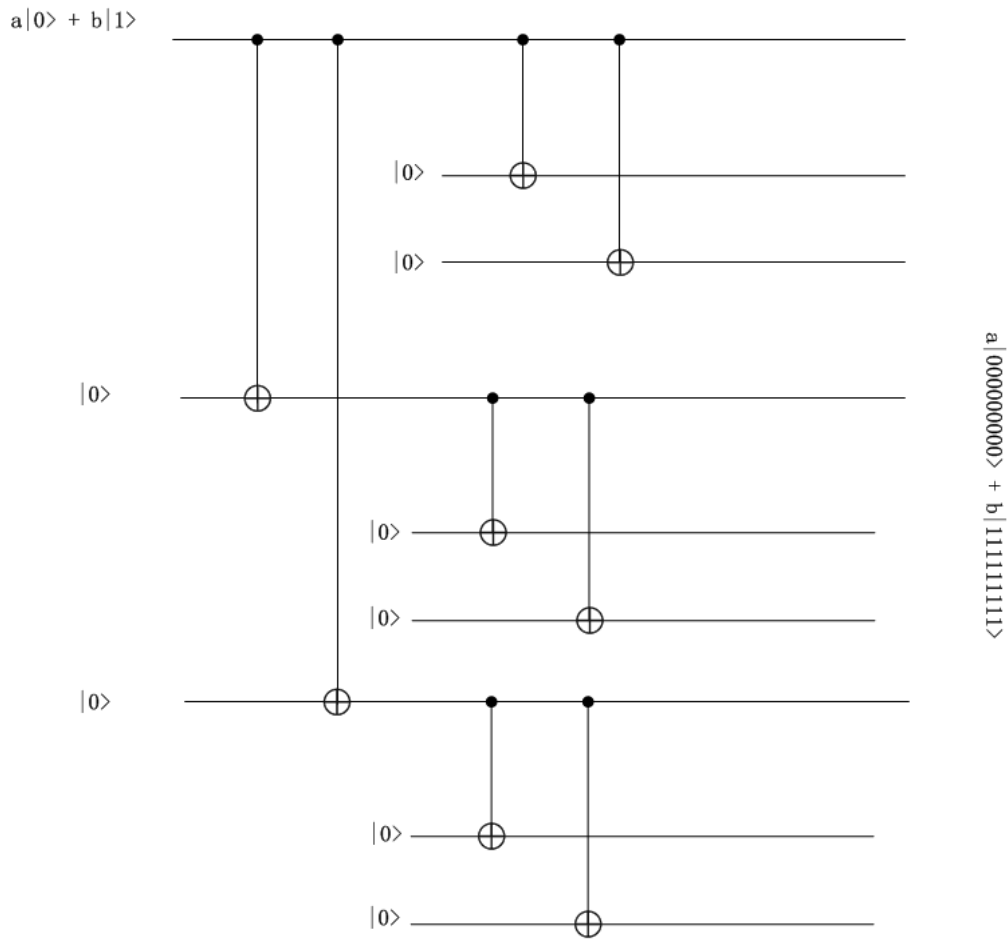

Fig 3. Quantum circuit for the 3rd level of multi-layer quantum secret sharing

For the purpose of threshold changeability, we show next that we can forget about the layers. In the original multi-layer QSS scheme, a member of the $n^{th}$ layer can further split the quantum secret to 3 more parties in the $(n+1)^{th}$ layer. Actually, we can see that every member can further split the secret to any more members, regardless of the layers. Since each member has a particle in the quantum entangled state:

$$a|\dots 0 \dots\rangle + b|\dots 1 \dots\rangle$$

By using the similar quantum circuit in Fig. 2 (with any number of auxiliary qubits $|0\rangle$ and corresponding Control-NOT gates), he can further split the quantum state to:

$$a|\dots 000 \dots 0 \dots\rangle + b|\dots 111 \dots 1 \dots\rangle$$

Then by sending these newly entangled qubits to more parties, the threshold has been increased.

We have shown how to share qubits with threshold changeability property. Now we briefly introduce how to use our TCSS construction to share classical bits. We can choose a random qubit basis such as:

$$\cos\left(\frac{\theta}{2}\right)|0\rangle \pm e^{i\varphi}\sin\left(\frac{\theta}{2}\right)|1\rangle$$

with random $0 \leq \theta \leq \pi, 0 \leq \varphi \leq 2\pi$. One of the basis vectors can be used for representing 1 and the other for 0. Then this qubit basis can be shared as above. Then after reconstructing, the qubit can be measured by this basis and recover the shared classical bit. In next section we will also show that this is a perfect threshold changeable SS scheme for classical information by choosing suitable parameters, i.e., any proper subset of the participating parties will gain no information about the secret classical bit.

## 4. Security Analysis

***Theorem 1.*** The quantum secret cannot be recovered if the participating number of parties is less than the current threshold $n$ (i.e., the number of all parties at this time).

***Proof:*** We show this by showing that if less than $n$ parties participate the reconstructing process, then the quantum states that they put together are mixed. While the original quantum state is clearly pure, hence the pure quantum secret state cannot be recovered.

Suppose the quantum secure pure state is $a|0\rangle + b|1\rangle$, and currently there are $n$ parties, then when all these parties put together their particles, they will get the following pure state:
$$a|00...0\rangle + b|11...1\rangle$$
This can be easily transformed to the quantum secret $a|0\rangle + b|1\rangle$. But if one or more parties are missing, then what the participating parties get will be a mixed one as shown next.

For simplicity, suppose there are 3 parties A, B and C, then when all 3 parties are present, they will get the pure state:
$$|\varphi\rangle_{ABC} = a|000\rangle + b|111\rangle$$
If one party is missing (w.l.o.g, suppose the last party C is missing), the remaining parties will get:
$$\rho_{AB} = Tr_C(|\varphi\rangle\langle\varphi|) = Tr_C(a|000\rangle + b|111\rangle)(a^*\langle000| + b^*\langle111|)$$
$$= Tr_C(|a|^2 |000\rangle\langle000| + ab^*|000\rangle\langle111| + ba^*|111\rangle\langle000| + |b|^2|111\rangle\langle111|)$$
$$= |a|^2|00\rangle\langle00| + |b|^2|11\rangle\langle11|$$

$$\rho_{AB} = \begin{pmatrix} |a|^2 & 0 & 0 & 0 \\ 0 & 0 & 0 & 0 \\ 0 & 0 & 0 & 0 \\ 0 & 0 & 0 & |b|^2 \end{pmatrix}$$

$$\rho_{AB}^2 = \begin{pmatrix} |a|^4 & 0 & 0 & 0 \\ 0 & 0 & 0 & 0 \\ 0 & 0 & 0 & 0 \\ 0 & 0 & 0 & |b|^4 \end{pmatrix}$$

Obviously, $Tr(\rho) = |a|^2 + |b|^2 = 1$ and $Tr(\rho^2) = |a|^4 + |b|^4 < 1$ if $a \neq 0$ and $b \neq 0$. So, the combined quantum state of A and B $\rho_{AB}$ is mixed.

If two parties are missing (w.l.o.g, suppose the first two parties A and B are missing):
$$\rho_C = Tr_{AB}(a|000\rangle + b|111\rangle)(a^*\langle000| + b^*\langle111|)$$
$$= Tr_{AB}(|a|^2 |000\rangle\langle000| + ab^*|000\rangle\langle111| + ba^*|111\rangle\langle000| + |b|^2|111\rangle\langle111|)$$
$$= |a|^2|0\rangle\langle0| + |b|^2|1\rangle\langle1|$$

$$\rho_C = \begin{pmatrix} |a|^2 & 0 \\ 0 & |b|^2 \end{pmatrix}$$

$$\rho_C^2 = \begin{pmatrix} |a|^4 & 0 \\ 0 & |b|^4 \end{pmatrix}$$

By similar argument above, we can see that the state of party C $\rho_C$ is also mixed.

This proves our theorem for three parties. By similar calculation, it is easy to prove that for any number of parties, the reconstructed state is mixed if not all parties are present. □

Note that our quantum secret share scheme is not perfect. It is well know that the Shamir's (t, n) secret sharing scheme is perfect. I.e., if less than *t* parties, no information about the secret will be revealed. In our scheme, partial information could be leaked. To quantify how much information could be leaked, next we calculate the fidelity between the partial state and the secret pure state. First, recall that the quantum fidelity [22] between two quantum state $\rho$ and $\sigma$ is defined as:

$$F(\rho, \sigma) \equiv \left(Tr\left(\sqrt{\sqrt{\rho}\sigma\sqrt{\rho}}\right)\right)^2$$

And if one the state, w.l.o.g. suppose it is $\sigma$, is pure, then:

$$F(\rho, \sigma) = \langle\sigma|\rho|\sigma\rangle$$

In our example above, $\rho_C$ is mixed as proved. And the quantum secret is pure $a|0\rangle + b|1\rangle$. So the fidelity between $\rho_C$ and the pure secret is:

$$(\bar{a}\ \bar{b})\begin{pmatrix}|a|^2 & 0 \\ 0 & |b|^2\end{pmatrix}\begin{pmatrix}a \\ b\end{pmatrix} = |a|^4 + |b|^4$$

And similarly the fidelity between $\rho_{AB}$ and the secret pure quantum state $a|00\rangle + b|11\rangle$ is also:

$$(\bar{a}\ 0\ 0\ \bar{b})\begin{pmatrix}|a|^2 & 0 & 0 & 0 \\ 0 & 0 & 0 & 0 \\ 0 & 0 & 0 & 0 \\ 0 & 0 & 0 & |b|^2\end{pmatrix}\begin{pmatrix}a \\ 0 \\ 0 \\ b\end{pmatrix} = |a|^4 + |b|^4$$

As can be seen from the example above, the fidelity for one party missing and two party missing are the same. And generally, the fidelities for any number of parties missing (from *1* to *n-1*) are the same. That just means more parties (under the provision that not all parties are present) will not help to find more information about the secret. This is related to the fact that any one party can easily further split and distribute the secret to any number of more parties. And each party (and even any subset of parties) hold basically the same partial information about the secret. Only when all the parties are present, the fidelity become 1, i.e., the secret can be reconstructed.

Of course, for security reason, we want the fidelity between the mixed partial quantum state and pure secret state should be as small as possible, i.e. we want as close to 0 as possible, since $0 \leq F \leq 1$. As shown above:

$$F = |a|^4 + |b|^4$$

We want to minimize the fidelity under the condition that $|a|^2 + |b|^2 = 1$. This can be done easily with basic calculus.

$$F = |a|^4 + |b|^4 = (|a|^2 + |b|^2)^2 - 2|a|^2|b|^2 = 1 - 2|a|^2(1 - |a|^2)$$

With $x = |a|^2$, the problem is then to maximize the second term to:

$$MAX\ y = 2x(1 - x)$$
$$S.T.: x \geq 0\ and\ x \leq 1$$

Differentiate the function we get $y' = 2 - 4x$. Let $y' = 0$, we get $x = 1/2$. So $y_{max} = 1/2$. I.e.,

$$F_{min} = 1 - \frac{1}{2} = \frac{1}{2}$$

This is the case where $|a|^2 = 1/2$ and $|b|^2 = 1/2$. I.e., the quantum secret state should be the following:

$$\frac{1}{\sqrt{2}}|0\rangle + e^{i\varphi}\frac{1}{\sqrt{2}}|1\rangle$$

Where $\varphi \in [0, 2\pi]$. I.e., for minimizing the fidelity we should use the quantum state on the equator of the Bloch sphere with polar angle $\theta = \pi/2$ and any azimuthal angle (See Fig 1.). Since a general qubit state can be represented as:

$$\cos\left(\frac{\theta}{2}\right)|0\rangle + e^{i\varphi}\sin\left(\frac{\theta}{2}\right)|1\rangle$$

If we want to use this scheme to share a classical bit 0 or 1, this just means that we should also use the basis on the equator of the Bloch sphere. For example, we can use $|+\rangle$ to represent 1 and $|-\rangle$ to represent 0. Then if all party are present, then the original $|+\rangle$ or $|-\rangle$ can be easily reconstructed, then measure it by the Pauli X operator, the classical bit can be easily recovered. Otherwise, if one or more parties are missing, then when measure the partial quantum state, the probability to get 0 and 1 are equal, just 1/2. Namely we have the following theorem.

***Theorem 2.*** Classical bit secret sharing based on our Q-TCSS scheme is perfect if using a basis on the equator of the Bloch sphere.

***Proof:*** Suppose we use

$$\frac{1}{\sqrt{2}}|0\rangle + e^{i\varphi}\frac{1}{\sqrt{2}}|1\rangle$$

to represent 1 and the corresponding orthogonal qubit state

$$\frac{1}{\sqrt{2}}|0\rangle - e^{i\varphi}\frac{1}{\sqrt{2}}|1\rangle$$

to represent 0 and there are three parties as above. Then if one party is missing, then the combined states of the other two parties are:

$$\rho_{AB} = \begin{pmatrix} |a|^2 & 0 & 0 & 0 \\ 0 & 0 & 0 & 0 \\ 0 & 0 & 0 & 0 \\ 0 & 0 & 0 & |b|^2 \end{pmatrix} = \begin{pmatrix} 1/2 & 0 & 0 & 0 \\ 0 & 0 & 0 & 0 \\ 0 & 0 & 0 & 0 \\ 0 & 0 & 0 & 1/2 \end{pmatrix}$$

Regardless of the sharing bits being 1 or 0. Hence the sharing bits is perfectly hiding when one party is missing. Similarly, if two parties are missing, the sharing bits is also hiding perfectly. And this proof works for not just three parties, it can easily be extended to any number of parties. So classical TCSS based on our quantum TCSS scheme is perfect. Which means no information is leaked if the participating parties are not all the members. □

Of course, any bases lying on the equator are equally good for a perfect classical secret sharing with threshold changeability. I.e., we can use the following basis for any $\omega$ (the $|\pm\rangle$ basis above is just the case $\omega = 0$).

$$\frac{1}{\sqrt{2}}(|0\rangle \pm e^{i\omega}|1\rangle)$$

Another example of $\omega = \frac{\pi}{10}$ is marked by dot on Fig 1.

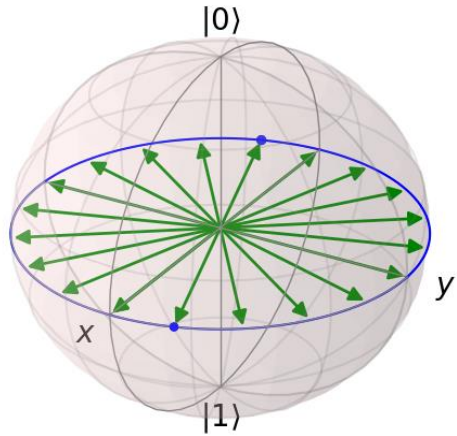

Fig 4. Pure quantum state on the equator of the Bloch sphere

## 5. Conclusions

In classical TCSS scheme, the changed threshold is not mandatory. Instead, the new threshold is negotiated among the reconstructing parties. So, when an adversary has corrupted the minimum number of parties, then the secret is revealed. The changeable threshold is void. Even if the parties are honest and delete old shares. Still an adversary or hacker may recover the deleted information from garbage or backup, and use them to reconstruct the secret.

We show in this paper that a quantum TCSS scheme can overcome the above loopholes of classical TCSS. In our Q-TCSS scheme, when the threshold has been changed, for example for 3 to 9, then all 9 parties should be presented in order to reconstruct the secret. Less than 9 parties cannot reconstruct the secret. We show that this kind of threshold changeability is almost impossible in classical computing, since classical information can be easily copied or be recovered from deletion. So, we show quantum advantage of TCSS scheme.

Although quantum information shared by our Q-TCSS scheme is not perfect, i.e., some partial information may be leaked if some (not all) parties try to reconstruct the quantum secret. Since the minimum fidelity we achieve is 1/2. We show classical bit shared by our Q-TCSS scheme can be perfect. I.e., no information about the classical bit is leaked when some (not all) parties try to steal the shared classical secret.

Our changed threshold is always the number of all the participating parties. This may not be flexible enough in some scenarios. In the future, we plan to construct $(t, n)$ Q-TCSS scheme, in which both the parameter $t$ and $n$ are changeable (of course we should always have $t > n/2$ according to the quantum no-cloning theorem).

**Reference**
[1]. Shamir. How to share a secret. Communications of the ACM, 22(11):612–613, 1979.
[2]. Asmuth, C., & Bloom, J. (1983). A modular approach to key safeguarding. IEEE Transactions on Information Theory, 29(2), 208–210.
[3]. Hsu, CF., Harn, L. Multipartite Secret Sharing Based on CRT. Wireless Pers Commun 78, 271–282 (2014). https://doi.org/10.1007/s11277-014-1751-x
[4]. Blakley, G. R. (1979). Safeguarding cryptographic keys. In Proceedings of AFIPs I979 national computer conference, New York (Vol. 48, pp. 313–317).

[5]. B. Chor, S. Goldwasser, S. Micali, and B. Awerbuch. Verifiable secret sharing and achieving simultaneity in the presence of faults. In Proceedings of the 26th IEEE Symposium on the Foundations of Computer Science (FOCS), pages 383–395, 1985.

[6]. Herzberg, A., Jarecki, S., Krawczyk, H., Yung, M. (1995). Proactive Secret Sharing Or: How to Cope With Perpetual Leakage. In: Coppersmith, D. (eds) Advances in Cryptology — CRYPT0' 95. CRYPTO 1995. Lecture Notes in Computer Science, vol 963. Springer, Berlin, Heidelberg. https://doi.org/10.1007/3-540-44750-4_27

[7]. Keju MENG, Fuyou MIAO, Yu NING, Wenchao HUANG, Yan XIONG, Chin-Chen CHANG. A proactive secret sharing scheme based on Chinese remainder theorem. Front. Comput. Sci., 2021, 15(2): 152801 https://doi.org/10.1007/s11704-019-9123-z

[8]. Joseph Y Halpern and Vanessa Teague. Rational secret sharing and multiparty computation: extended abstract. In Symposium on the theory of computing, 2004, pages 623–632, 2004.

[9]. S Dov Gordon and Jonathan Katz. Rational secret sharing, revisited. In Security and cryptography for networks, 2006, pages 229–241, 2006.

[10]. Hillery, M., Buzek, V. & Berthiaume, A. Quantum secret sharing. Phys. Rev. A 59, 1829 (1999).

[11]. Cleve, R., Gottesman, D. & Lo, H.-K. How to share a quantum secret. Phys. Rev. Lett. 83, 648 (1999)

[12]. Zhang, Kj., Zhang, X., Jia, Hy. et al. A new n-party quantum secret sharing model based on multiparty entangled states. Quantum Inf Process 18, 81 (2019). https://doi.org/10.1007/s11128-019-2201-1

[13]. Cheng X , Guo R , Chen Y , et al. Improvement of a multi-layer quantum secret sharing based on GHZ state and Bell measurement[J]. International Journal of Quantum Information, 2018, 16(06).

[14]. Kaushik Senthoor and Pradeep Kiran Sarvepalli. Communication efficient quantum secret sharing. Phys. Rev. A, 100:052313, Nov 2019.

[15]. Maitra A, De S J, Paul G, et al. Proposal for quantum rational secret sharing. Phys. rev. a, 2015, 92(2): 022305

[16]. Qin, H., Tang, W.K.S. & Tso, R. Rational quantum secret sharing. Sci Rep 8, 11115 (2018). https://doi.org/10.1038/s41598-018-29051-z

[17]. Victoria Lipinska, Gl´aucia Murta, J´er´emy Ribeiro, and Stephanie Wehner. Verifiable hybrid secret sharing with few qubits. Phys. Rev. A, 101:032332,Mar 2020.

[18]. Martin K, Pieprzyk J, Safavi-Naini R, Wang H. Changing thresholds in the absence of secure channels. In: Proceedings of Springer Australasian Conference on Information Security and Privacy. 1999, 177–191

[19]. Steinfeld R, Wang H, Pieprzyk J. Lattice-based threshold changeability for standard Shamir secret-sharing schemes. IEEE Transactions on Information Theory, 2007, 53(7): 2542–2559

[20]. Meng K, Miao F, Huang W, Xiong Y. Threshold changeable secret sharing with secure secret reconstruction. Information Processing Letters, 2020, 157: 105928

[21]. Harn, L., Hsu, C. & Xia, Z. A novel threshold changeable secret sharing scheme. Front. Comput. Sci. 16, 161807 (2022). https://doi.org/10.1007/s11704-020-0300-x

[22]. Nielsen M A, Chuang I L. Quantum Computation and Quantum Information: 10th Anniversary Edition[M]. Cambridge University Press, 2011.